# When Social Sensing Meets Edge Computing: Vision and Challenges

Daniel (Yue) Zhang, Nathan Vance, Dong Wang
Department of Computer Science and Engineering
University of Notre Dame, IN, USA
Email: {yzhang40, nvance1, dwang5}@nd.edu

*Abstract*—This paper overviews the state of the art, research challenges, and future opportunities in an emerging research direction: Social Sensing based Edge Computing (SSEC). Social sensing has emerged as a new sensing application paradigm where measurements about the physical world are collected from humans or from devices on their behalf. The advent of edge computing pushes the frontier of computation, service, and data along the cloud-to-things continuum. The merging of these two technical trends generates a set of new research challenges that need to be addressed. In this paper, we first define the new SSEC paradigm that is motivated by a few underlying technology trends. We then present a few representative real-world case studies of SSEC applications and several key research challenges that exist in those applications. Finally, we envision a few exciting research directions in future SSEC. We hope this paper will stimulate discussions of this emerging research direction in the community.

## I. INTRODUCTION

Social sensing has become a new sensing paradigm for collecting real-time measurements about the physical world from humans or mobile devices on their behalf [1]–[5]. Examples of social sensing applications include urban traffic monitoring using mobile apps [6], obtaining real-time situation awareness in the aftermath of a disaster using self-reported observations from citizens [7], and smart healthcare monitoring using wearable sensors [8]. A key limitation in the current social sensing solution space is that data processing and analytic tasks are often done on a "backend" system (e.g., on dedicated servers or commercial clouds) [2], [9]–[11]. Unfortunately, this scheme ignores the rich processing capability of increasingly powerful edge devices owned by individuals (e.g., mobile phones, tablets, smart wearables, and the Internet of Things). For example, the emerging AI accelerators (commonly called "AI Chip") on smartphones are capable of finishing complex deep learning tasks that are traditionally done on large server racks [12]. These *ubiquitous, powerful, and individually owned* devices are referred to as "edge devices" in this paper.

The advent of edge computing pushes the frontier of computation, service, and data along the cloud-to-things continuum to the edge of the network [13]–[15], and brings new opportunities for social sensing applications. By combining social sensing with edge computing, the privately owned edge devices not only serve as pervasive sensors, but also form a federation of computational nodes where the data collected from them can be processed and consumed at the edge [16]–[19]. We refer to the marriage of social sensing and edge computing as Social Sensing based Edge Computing paradigm, or SSEC for short. We illustrate a typical SSEC system architecture in Figure 1. The SSEC system consists of an edge layer, an edge server layer, and a service layer. In the edge layer, privately owned edge devices (e.g., mobile phones, IoT devices, drones) are leveraged to perform the sensing, storage, networking, and computational tasks near the source of the data. The edge server layer[1] (often comprised of local servers, cloudlets, smart routers, or gateways) provides an intermediate layer between the edge devices and the cloud. The edge server layer also provides additional data storage and computing power in locations of close proximity to the edge devices [20], [21]. The service layer (often built into a back-end cloud) provides a global service interface to all users interested in the applications/services.

The advantages of the SSEC paradigm are multi-fold: 1) social sensing applications can process the sensing data right at the edge devices where the data has been collected, which could significantly reduce the communication costs (e.g., bandwidth) and improve the Quality of Service (QoS) (e.g., delay) of the applications; 2) social sensors (e.g., owner of the edge devices) can obtain payoffs/rewards by leveraging the idle resources of their devices to execute the computational tasks for the application; 3) the SSEC architecture does not suffer from a single point of failure and alleviates the performance bottleneck of the "back-end" solutions.

The SSEC paradigm also introduces many research challenges. In particular, SSEC introduces a set of new challenges to real-time resource management by supporting delay-sensitive social sensing applications in edge computing systems. For example, the edge devices (often owned by end users) in SSEC are generally opportunistic and selfish (e.g., they are not committed to or interested in executing the sensing tasks or sharing their private device status unless incentives/payoffs are provided) [22], [23]. This assumption is unique in SSEC and contrasts sharply with the assumption made in the "backend" based solutions in traditional distributed or cloud-based systems where all computational devices are fully committed and information is shared among all devices [10]. Furthermore, the SSEC paradigm calls for close collaboration among end users, infrastructure owners, and application managers. Due to the lack of natural trust among them, none of these parties can be fully trusted as they might be interested in performing privacy and security attacks.

---

[1]The edge server is also commonly referred to as a *fog node* in fog computing literature. We use these two terms interchangeably in this paper.

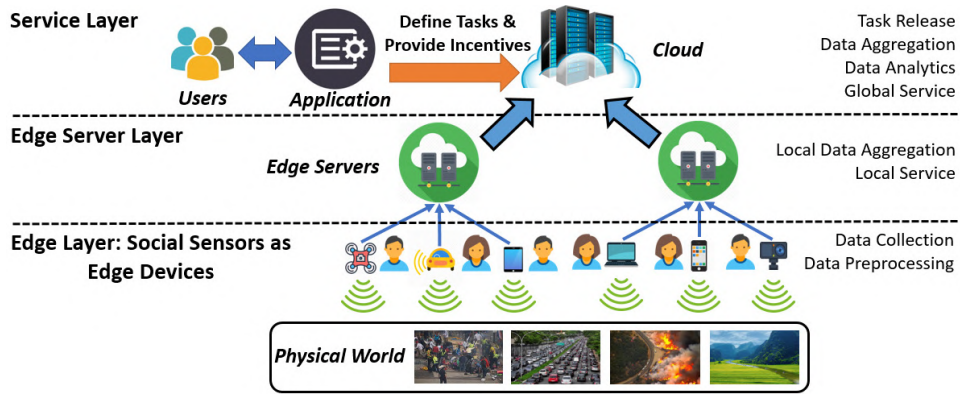

Figure 1: Social Sensing based Edge Computing

The rest of the paper is organized as follows. In Section II, we present the definition, enabling technologies, and impact of the SSEC paradigm. In Section III, we discuss several important applications and case studies of SSEC. In Section IV, we discuss the unique research challenges and opportunities in SSEC. We outline the future road map of this direction in Section V. Finally, we conclude our vision of SSEC in Section VI.

## II. Social Sensing Edge Computing Paradigm

IoT devices owned by individuals are increasingly equipped with powerful computing and diverse sensing capabilities. The sensing data generated by these devices provides an alternative lens into physical phenomena as compared to traditional sensor networks [1], [24]–[26]. Due to the sheer volume of data generated by these devices, it makes sense to explore opportunities for processing the data at the edge of the network [10]. Previous work in edge computing leverage cloudlets [20], micro datacenters [27], and fog computing [28] to address the deficiency of cloud computing when the data is produced at the edge of the network. However, these solutions fail to take advantage of privately owned edge devices as SSEC does, and they instead rely on infrastructure which must be provisioned ahead of time. In this section, we formally define social sensing based edge computing (SSEC) and discuss how SSEC is complementary to existing edge computing frameworks.

### A. What is SSEC?

**DEFINITION 1. Social Sensing based Edge Computing (SSEC)**: an application paradigm that uses humans and devices on their behalf to sense, process, and analyze data collected about the physical world.

In this definition, the devices owned by individuals not only collect data about the physical world, but also actively participate in the application by performing computations and analytic tasks. These privately owned edge devices can be quite heterogeneous, ranging from a GPS sensor, Raspberry Pi, or robot, to a powerful multi-processor server.

SSEC has two important features: 1) it is human-centric; and 2) it has the flexibility to support various applications with different system architectures. We elaborate on these features below.

*1) Human-centric Nature of SSEC:* SSEC is human-centric. On one hand, the owners of the edge devices are freelance users and their unique concerns must be carefully considered in the SSEC paradigm. These human concerns includes privacy and security, compliance and churn, and incentives, which will be elaborated in Section IV. On the other hand, we envision that not only can devices engage in the sensing and computational tasks, but people can directly participate as well. In fact, many social sensing applications require input directly from a human, such as reporting traffic congestion [29], or taking videos of an emergency event [18]. Also, SSEC considers the potential of people serving as "social edge nodes" where they directly make inferences using the data. For example, consider an abnormal event detection scenario where edge devices are used to collect video data and infer abnormal events such as an intrusion [30]. Instead of using machine learning algorithms to perform such data analytic tasks, humans can directly identify the abnormal events from the video with high accuracy [31]. We explore the possibility of leveraging humans as computing nodes in a pioneer work [7]. This unique feature of SSEC where human input and intelligence complements the existing edge/cloud computing paradigm promises to enable new applications that would not be possible without it.

*2) Flexibility of SSEC to Support System Variations:* Like traditional edge computing systems that come in many different architectures [32], SSEC has diverse system variations as well (Figure 2). While SSEC focuses more on the privately owned edge devices, it by no means intends to drastically replace the existing cloud or edge computing paradigm by diminishing the existing infrastructure such as cloud servers, large data centers, cloudlets, or near-edge micro data centers. In fact, SSEC fully takes advantage of existing system infrastructures. The choice of the system architecture is application specific. We summarize a few representative architectures below.

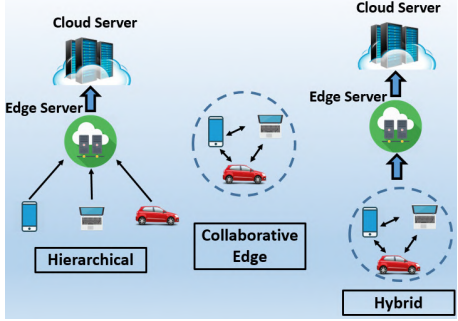

Figure 2: Example SSEC System Variations

**Hierarchical:** A typical cloud-edge hierarchical SSEC system architecture is shown in Figure 2. It follows the hierarchical structure where a remote cloud server, which is often powerful and has a massive storage capacity, manages the application and provides a global interface to the users. The application governs a set of spatially distributed edge clusters, where an edge cluster consists of a local edge server (e.g., a micro data center or a Road-Side-Unit) and the nearby edge devices that connect to it. In [33], [34], typical edge clusters are illustrated, including a set of devices in a coffee shop connected to a small in-house server owned by the shop; a set of vehicles connected to a Road-side-Unit (RSU) on the same street; and a set of mobile phones connected to the nearest base station. The key characteristic of this hierarchical structure is that the data flow is static: edge devices process the data locally and offload further computational tasks to the edge servers, and edge servers further process the data and send the results to the cloud server for data aggregation tasks and storage.

**Collaborative Edge:** In a collaborative edge architecture, edge devices in close proximity self-organize into a computing cluster and provide peer-to-peer services such as content delivery and computation offloading. This architecture is particularly suitable for application scenarios where edge or cloud servers are not readily available, or to avoid periodic costs by using these infrastructures. Consider a crowd video sharing application example where a set of spectators at a sporting event (e.g., a soccer game) can take videos of the highlights of the game and can stream them to people in the audience who missed the play or who sit in a undesirable locations. To improve performance for devices with poor network connections, the system can encode the video streams to a lower bitrate. In such a scenario, a remote cloud can introduce significant delay for video sharing and local edge servers may not be available (the servers/smart gateways at the stadium may not be accessible by the audience). Therefore, in the collaborative edge architecture, privately owned edge devices perform these typically server-side roles.

**Hybrid:** A hybrid system architecture is a combination of both a hierarchical and collaborative edge, in which self-organized edge devices are connected to the available infrastructure (i.e., edge servers and the cloud). This infrastructure is ideal for scenarios where self-organized edge devices cannot satisfy QoS requirements, so readily available edge servers and the cloud are leveraged to boost performance. Consider a disaster response application where edge devices collaboratively report damages during a disaster, often by executing image analysis and machine learning algorithms to classify damage severity [7]. A computationally weak edge device such as a video camera can collect image data of the affected area and offload the damage assessment task to a powerful edge device nearby via Bluetooth. The assessment result is further reported to all nearby edge devices in the form of alerts. In the case where edge devices are under-performing due to lack of high-end processors, the collaborative edge can offload tasks to nearby edge servers, such as base stations, or cloud servers for further processing.

### B. Why We Need SSEC

Social Sensing based Edge Computing (SSEC) is motivated by a few key technical trends: i) the IoT devices owned by individuals are becoming increasingly powerful and some of them even have similar computing power as the dedicated servers in traditional edge computing systems [18], [35]. Therefore, it becomes a growing trend to push the computation to the edge devices rather than dedicated remote servers or edge servers [33]; ii) the popularity of mobile payments provides a more convenient way for individuals to receive incentives by contributing the spare resources on their IoT devices for accomplishing social sensing tasks [36]. We summarize a few advantages of SSEC below.

*1) Coverage and Availability:* One of SSEC's main advantages is its coverage and the availability of edge devices. There are billions of privately owned edge devices worldwide that can collect and process data at a global scale. This natural mobile network is clearly advantageous in terms of coverage as compared to static infrastructure such as data centers or surveillance cameras. Furthermore, SSEC provides mobility as the sensing and computing resources move geographically with their users. This makes SSEC ideal for people-centric sensing and computing tasks as the availability of resources is closely correlated with the prevalence of noteworthy events.

*2) Delay Reduction:* Social sensing applications can process the sensing data on the edge devices where the data has been collected or on devices in close proximity, which could significantly reduce the communication costs (e.g., bandwidth) and improve the Quality of Service (QoS) (e.g., delay) of the applications. This makes SSEC ideal for real-time or time-sensitive applications.

*3) Utilization:* SSEC fully leverages the sensing and computing power of the edge devices. Compared to traditional edge computing frameworks that offload computational tasks to edge servers or cloud servers, SSEC envisions that tasks can be executed on smart devices owned by individuals as well. By pushing the tasks to the edge, the SSEC architecture removes the single point of failure and alleviates the performance bottleneck of the "back-end" solution. This enables SSEC to avoid high deployment costs for sensing tasks, and to save money on the back-end infrastructure.

*4) Reward Earnings:* In SSEC, participants can obtain rewards by contributing the idle resources of their devices to execute computing tasks for the SSEC application. Similar to how unused compute cycles are sold in cloud environments, this creates a new market where the idle resources of edge devices can now be fully utilized.

## III. REAL-WORLD APPLICATIONS

In this section, we discuss a few representative SSEC applications in real world scenarios.

### A. Disaster and Emergency Response

An important application of SSEC is to provide real-time situation awareness during disaster and emergency events (e.g., forest fire, robbery, terrorist attacks) [37], [38]. During such events, human sensors (e.g., citizens, first responders, news reporters) often spontaneously report a massive amount of sensing information that describes the unfolding of the event. SSEC provides a suitable architecture for this category of applications: 1) the edge devices, with close proximity to the human sensors, can collect and extract useful features about the event without sending all data streams back to the cloud; 2) the edge server layer in SSEC can gather processed data and exacted features from edge devices to provide real-time event updates for local citizens; 3) the cloud server aggregates all information collected and provides it to relevant agencies and/or the general public. Figure 3 illustrates a scenario where people use mobile phones and cameras to provide first-hand footage of a terrorist attack at a shopping center. These data can be helpful in tracking a suspect's escape path. The edge server layer provides time-critical alerts for potential threats and offers safety recommendations.

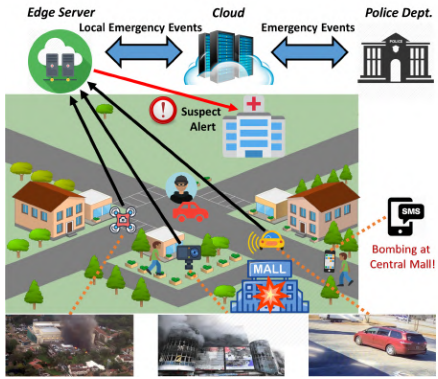

Figure 3: Disaster and Emergency Response Application

### B. Collaborative Traffic Monitoring

Collaborative traffic monitoring in social sensing aims at collecting timely information about traffic conditions (e.g., congestion, accidents, and events) of an area of interest (e.g., a city). Such applications are useful for many transportation services such as route planning, traffic management, and fuel efficient navigation [34]. Traditionally, traffic monitoring has been performed by analyzing data from statically installed traffic cameras, which suffers from poor coverage [39]. Moreover, the data generated by these traffic cameras were processed at a remote cloud server, which introduces significant delay and bandwidth costs. SSEC can address this problem by fully leveraging social sensing and the edge devices owned by people. In particular, the personally owned sensing devices on vehicles (e.g., cameras, accelerometers, GPS sensors) offer opportunities to collect a large amount of traffic data in real time. For example, a typical traffic monitoring application can task a set of drivers to use their dashboard cameras to record traffic in front of their vehicles. The processed data (e.g., extracted features) are offloaded to nearby edge servers (i.e., RSU) for further analysis of traffic conditions. Additionally, human sensors are also capable of reporting high-level descriptions of the traffic context using their smart phones. An example of such a social sensing application is Waze[2] where drivers collectively report their observations of accidents, road hazards, and traffic jams in real-time. In Figure 4, pedestrians and drivers collaboratively contribute traffic data using their edge devices. The edge server infers the traffic conditions of local streets from the social sensing data and sends accident alerts to the drivers. Transportation agencies can also query the cloud for the road conditions and accidents in their regions of jurisdiction and prioritize accident response, road repair, or traffic control accordingly.

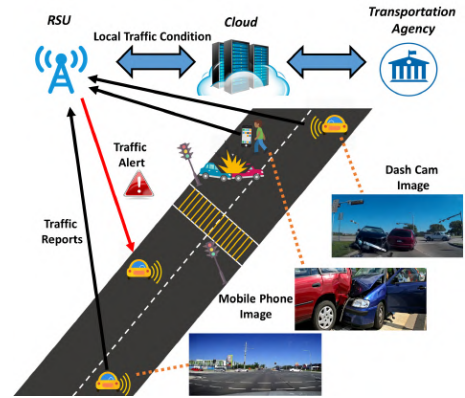

Figure 4: Collaborative Traffic Monitoring Application

### C. Crowd Abnormal Event Detection

The goal of crowd abnormal event detection in social sensing is to generate alerts for abnormal events from data contributed by human sensors and their portable devices (e.g., mobile phones). Traditional abnormal event detection solutions largely depend on video data collected from installed surveillance cameras and utilize image processing techniques to identify these events [40], [41]. Those solutions fail in situations where installed cameras are not available (e.g., due to deployment costs). The prevalence of camera-enabled portable devices has enabled the collection of geo-tagged pictures, videos, and user-reported textual data through social sensing applications. Such multi-modal data can be exploited for enhanced situation awareness during abnormal activities (e.g., providing insights for investigating the severity and causes of events). For example, during a soccer game, events such as sudden appearance of unexpected object or malicious behavior of people (e.g., throwing a signal flare into the field) can pose great threats to the safety of players and interrupt the normal course of the game (Figure 5). In our SSEC framework, the audience (as human sensors) can contribute videos, images, and texts to report their observations about the abnormal events. Upon detection of the abnormal events during the game, the cloud-hosted service will send alerts to the fans and the police department for an emergency response.

---

[2]https://www.waze.com/

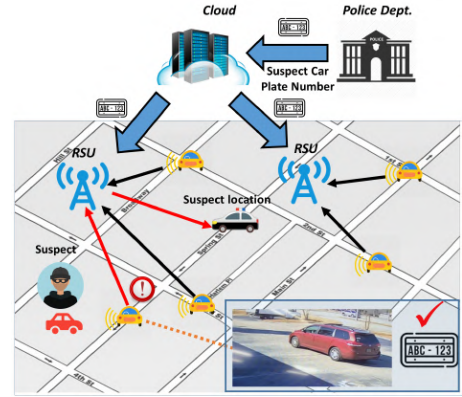

Figure 6: Plate Recognition Application

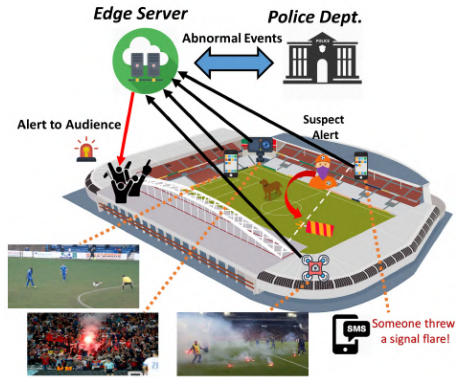

Figure 5: Crowd Abnormal Event Detection Application

### D. Plate Recognition

The plate recognition application (Figure 6) was first introduced in an effort to leverage private vehicles to collaboratively track down suspects of AMBER alerts [42]. In this application, vehicles equipped with dash cameras form a city-wide video surveillance network that tracks moving vehicles using the automatic license plate recognition (ALPR) technique. This system can be used to effectively track down criminal suspects who are on the run in vehicles. It complements existing vehicle searching processes that heavily rely on reports from witnesses who might miss alerts and cannot search enough areas of city [43]. Collecting surveillance video footage can expand coverage. However, analyzing huge amounts of video data in the cloud leads to unreasonable data transmission costs and high response latency. SSEC can significantly reduce the cost of data transmission and response latency by offloading the data to nearby RSUs for real-time processing. SSEC also pushes local processing to be done on these private vehicles to extract features from the raw images and send the processed data to the RSUs instead. This is because the video data collected from the vehicles can also reveal private information of the drivers (e.g., residence location) or the faces of the citizens. Upon detecting the suspect's vehicle, the cloud-hosted service will send alerts to the police department for an immediate response.

### E. Crowd Video Sharing

The crowd video sharing application (Figure 7) uses self-organized edge devices to perform peer-to-peer video content delivery. This application is most suitable for events where people take interesting videos and want to share it with one another. For example, if a spectator at a soccer match has a good view of some action, then other spectators in less favorable locations may desire to view the footage from the better perspective. In order to facilitate this application the system must 1) employ the participating edge computing resources to avoid bottlenecks as the system scales, and 2) perform video encoding so that devices with poor network connections can be sent smaller video files, thus avoiding network delays. This problem can be solved using SSEC by coordinating edge devices to perform computation and communication tasks, thus providing a source of compute power and bandwidth which scales with the number of participating devices, i.e., demand. A bottom-up game theoretic decision making process optimizes the encoding and transmission of the videos in order to minimize delay in the system [44].

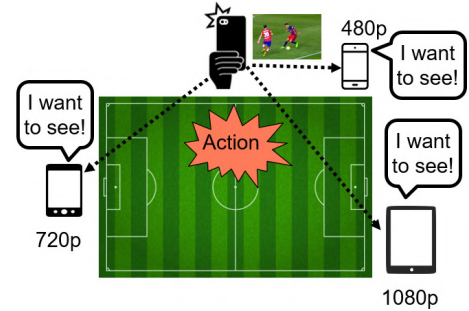

Figure 7: Crowd Video Sharing Application

## IV. RESEARCH CHALLENGES AND OPPORTUNITIES

The fusion of social sensing and edge computing pushes the frontier of sensing, computation, and service to the edge of the network where social sensing occurs. However, utilizing edge devices in the context of social sensing introduces a set of fundamental challenges that are yet to be fully addressed. In this section, we discuss a few critical research challenges and opportunities in SSEC.

### A. Resource Management with Rational Edge

In SSEC, the edge devices are usually owned by end users rather than application providers. Due to the rational nature of device owners, edge devices and applications often have inconsistent or even conflicting objectives [16]. We refer to this unique feature of SSEC as *"rational edge"*. Due to the rational edge feature, two important issues prevent existing resource managements schemes from being applied to SSEC, namely *competing objectives* and *asymmetric information*.

*1) Competing objectives:* From the application's perspective, it is important to ensure that the edge devices finish the allocated social sensing tasks in a timely fashion to meet the Quality of Service (QoS) requirements (e.g., end-to-end delays). In contrast, device owners are often less concerned about the QoS of the applications but are instead concerned about their costs in running the computational tasks allocated by the applications (e.g., the device's current utilization, energy consumption, memory usage). Thus, they are often unwilling to execute the allocated tasks until sufficient incentives are provided [22]. This is in sharp contrast with traditional distributed computing systems where computational resources are fully cooperative and directly controlled by the application. The mismatch in objectives held by the end users and the application must be carefully addressed by developing a set of new computation allocation models that respect such discrepancies between the two parties.

*2) Asymmetric Information:* Another critical challenge in SSEC is that the application server and edge devices usually have different degrees of information, i.e., *"asymmetric information"*. Such asymmetric information makes resource management in social sensing based edge computing systems particularly challenging [16]. The asymmetric information challenge can be viewed from two aspects. On the server side, the application normally has detailed information about the tasks (e.g., the dependencies and criticality of the tasks). This information is important in understanding how tasks are related to the QoS requirements imposed by the social sensing application (e.g., which tasks are more important and should be prioritized; which tasks should have a tighter deadline). In contrast, the edge devices are often less concerned about the details of the tasks and the servers' QoS requirements but more interested in their own device status (e.g., CPU utilization, energy consumption, memory usage). Moreover, an edge device may not share its status information with the server or other edge devices in the system due to various concerns (e.g., privacy, energy, bandwidth). This leads to insufficient information for the server to make optimal computation allocation decisions.

## B. Constrained Cooperativeness

In SSEC, edge devices are assumed to be only partially cooperative in finishing their computational tasks due to the rational or selfish nature of end users. This challenge is referred to as *"Constrained Cooperativeness"*. Previous studies showed that collaboration among computation nodes can significantly improve efficiency of resource utilization in distributed systems [45]. Such collaboration between edge devices in social sensing applications is essential to achieve optimized scalability and efficiency in the SSEC system. For example, the execution time of a set of tasks can be significantly reduced if those tasks are allocated to a group of edge devices that run the tasks in parallel and finish them collaboratively. Consider an abnormal event detection application where edge devices (e.g., smartphones, dash cameras) are tasked to take videos or pictures of their surroundings to detect abnormal activities. An edge device may not be equipped with a camera and thus is incapable of completing the allocated tasks on its own. On the other hand, if it has strong computing power, then it can serve as a "local computation hub" for nearby lower-end edge devices that do have cameras. However, collaboration among edge devices is especially challenging because: i) edge devices are rational actors who are unwilling to collaborate with others unless sufficient incentives are provided; ii) various constraints may prohibit collaboration among edge devices (e.g., latency constraints imposed by the physical distance between devices or trust constraints imposed by the trust between devices); iii) collaboration requires explicit consideration of the *task dependencies* of the application.

## C. Pronounced Heterogeneity

The heterogeneity in SSEC is often more pronounced than in regular edge computing systems. In particular, the edge devices in SSEC often have diversified computing power, runtime environments, network interfaces, and architectures, making it difficult to orchestrate these devices to collaboratively accomplish the sensing and computational tasks. The heterogeneity problem in SSEC is particularly challenging because it is not possible for the application to cherry-pick the devices in a fully controlled manner given the fact the devices are owned by individuals [46]. In order to tame the heterogeneity of edge devices in SSEC, several critical research tasks are involved.

*1) Runtime Abstraction:* A critical issue in heterogeneous SSEC is that the devices have diverse runtime environments that may not support the social sensing tasks to be processed. For example, a device may have an incompatible operating system or lack the necessary dependencies to execute a social sensing algorithm (e.g., a deep learning algorithm cannot run on a device without necessary libraries such as Tensorflow or CUDA [47]). Containerization techniques such as Docker [48] can abstract away some hardware details of the devices and provides a virtual environment that offers a lightweight, portable and high-performance sandbox to host various applications. In particular, the social sensing application developers can "wrap" all necessary dependencies and the OS itself into a Docker container for each social sensing application. Such runtime abstractions can allow the edge devices in SSEC to provide the same interface to the social sensing application developers and offers them the "write once and run anywhere" feature despite the heterogeneity of SSEC devices.

*2) Hardware Abstraction:* Hardware abstraction targets at abstracting away the details of heterogeneous hardware specifications of the edge devices for the ease of resource management in SSEC. A possible solution was proposed in HeteroEdge [17], where the hardware capabilities of a device can be represented as a set of "workers". HeteroEdge considers three types of workers that are essential for finishing social sensing tasks in SSEC - CPU, GPU, and Sensor workers. Each worker is associated with a capability descriptor in terms of the estimated worst case execution time (WCET) for processing social sensing tasks. The device owners can specify which workers are available to the SSEC application. HeteroEdge follows three important design principles in hardware abstraction in SSEC: i) the set of heterogeneous edge devices should form a unified homogeneous resource pool for the social sensing application; ii) the device owners should be able to control which resources they would like to provide for an application; iii) the edge devices can easily keep track of their own dynamic status and provide necessary context information for the runtime decision and optimization in SSEC.

*3) Networking Abstraction:* The privately owned edge devices in SSEC can have very heterogeneous network interfaces (e.g., Bluetooth, WiFi, Zigbee) and it is essential to abstract away the networking details to allow developers to deploy SSEC applications without worrying about the specific network interface and protocol. A promising technique to accomplish this task is Software Defined Networking (SDN) [49]. SDN can orchestrate the network, the services, and the devices by hiding the complexities of this heterogeneous network environment from the end users. It provides APIs that can simplify the management of the network, define network flows, and facilitate virtualization within the network.

We found existing resource management work in edge computing cannot sufficiently handle the pronounced heterogeneity in SSEC. A middleware that jointly addresses the three levels of abstraction above for SSEC has yet to be developed.

### D. Robustness against Churn and Dynamic Context

In SSEC, edge devices are most often privately owned and managed, and therefore suffer from churn [21], causing inconsistent availability by devices in edge computing. The inconsistency of edge device availability is aggravated since devices routinely kill tasks for power savings, or are opportunistically contributing compute power and then must stop in order to service their primary purpose [50]. Furthermore, in the case of mobile computing systems, a main criterion in the eligibility of a device to perform a task is the location of that device. Should the device move, then it may become unable to serve its function and must be replaced by a device in a more favorable location. To solve this problem in a way that is both scalable and reliable, we introduce buffering into multi-stage streaming applications. In such systems, tasks are broken into multiple stages where different devices perform an operation at each stage of a computational pipeline. If a device along the pipeline unexpectedly quits and must be replaced, then the replacement can be "filled in" by the the devices adjacent to it in the pipeline. Furthermore, this pipeline design lends itself to taking fine-grained advantage of heterogeneous edge computing hardware since each stage can be matched to a specialized computing platform.

Another challenging issue in the SSEC system is that edge devices have volatile statuses and their willingness to participate in SSEC applications may change dynamically over time. We refer to this challenge as *dynamic context*. Consider an environment sensing application where edge devices (e.g., mobile phones) are used to collectively monitor the air pollution of a city. Each edge device is tasked to monitor a particular area. An edge device (or its owner) may change the compliance of task execution due to i) changes in the battery status of the device, or ii) changes in the physical location of the device with respect to the monitored location. Failure to capture such dynamics may lead to significantly suboptimal resource allocation where the costs of edge devices to complete a task are prohibitively high.

### E. Privacy and Security

SSEC entails potential privacy risks to owners of edge devices in social sensing applications. During the data collection phase, the data collected from edge devices can potentially reveal end users' private information. For example, in the plate recognition application, the image captured by an edge device may contain street information, potentially disclosing user residence or mobility patterns. Similarly, during the resource management phase it is of the application's interest to obtain better knowledge on the status of each edge device to maximize the task allocation efficiency. However, the edge devices may not be willing to share such status information due to their privacy configurations. Existing privacy preserving techniques, such as anonymity techniques, can effectively protect the identities of edge devices from curious entities. But such techniques also prevent the application from identifying the contributors of the computational tasks, thereby preventing the server from distributing incentives through conventional means [34]. Though privacy-aware SSEC systems have been proposed [19], work still must be done to ensure that both the privacy expectations from end users and the QoS requirements from the applications are met.

Security in SSEC is an important concern, both for the benefit of users participating in the application and for the application itself so that the services rendered are not sabotaged. Unfortunately, the architecture of SSEC in which data originates and is processed on privately owned edge devices does not lend itself to conventional security systems (e.g., authentication in order to access a resource). Care must be taken, therefore, to ensure that i) peer-to-peer APIs in Collaborative Edge or Hybrid SSEC architectures are designed such that private information cannot be stolen by malicious attackers; ii) it is difficult or impossible to "game" the system by contributing incorrect sensing measurements or computation results in order to obtain the incentives without expending effort; iii) the system is resilient against attempts to sabotage or "poison" the results of the application for malicious purposes; iv) the system is resilient against denial-of-service attacks such as intentionally delaying tasks in order to harm the QoS.

## V. ROADMAP FOR FUTURE WORK

### A. SSEC and 5G

5G promises to have an estimated network speed as fast as 10 Gb/s and a network latency as low as 1 ms [51]. We envision that 5G will significantly boost the performance of SSEC and enable new SSEC applications. For example, the emergence of 5G networking capabilities will increase the number of connected devices on a network and promote collaboration among private edge devices. The delay requirement of 5G requires base stations to be deployed at a high density, which would also be able to serve as edge servers in SSEC. With 5G networks, SSEC applications that involve video content transfer such as crowd video sharing and plate recognition will significantly benefit from the boosted Internet speed and ultra low latency. We envision that more data intensive and delay sensitive SSEC applications will be enabled by 5G and future networking technologies.

### B. SSEC and AI

AI at the edge is a growing trend in both industry and academic research. Many AI-enabled chips have been developed and integrated into video cameras, hand-held devices, and vehicles [42]. However, AI capabilities are still far from being pervasive - many edge devices are low-end sensors without processing capabilities or hardware (e.g., GPU) for supporting AI algorithms. SSEC can promote AI by developing a collaborative intelligent edge where lower end devices can offload AI tasks to devices with AI capabilities. Many roadblocks must be removed for this vision to be realized. For example, performing AI tasks on privately owned edge devices inevitably incurs an energy cost. Considering that the battery is often the most precious resource of an edge device [14], incentive mechanisms must be designed so that a fair market can form to reward those who contribute energy. The collaborative intelligent edge also involves interactions among devices of differing ownership. Therefore, privacy and trust concerns must be carefully addressed.

### C. SSEC and Human-in-the-loop

Human-in-the-loop SSEC enables the integration of human intelligence (e.g., context-awareness, cognitive skills) with the processing and sensing capability of physical devices. We envision that the human component of SSEC can be modeled as a "social edge node" in which the human can perform inference or make decisions in the edge computing framework just like a physical device. This Human-in-the-loop SSEC paradigm can benefit many mission critical tasks by introducing the domain expertise that people possess. For example, humans can improve the effectiveness of physical systems in many intelligent tasks (e.g., disaster assessment [7] and traffic abnormality detection [6]). We envision that a new set of theories for building human-machine hybrid systems must be developed to fully leverage human intelligence in SSEC.

## VI. CONCLUSION

In this paper, we present an emerging SSEC framework to exploit the edge-enabled infrastructure and the ever-increasingly powerful IoT devices to improve the scalability and responsiveness of social sensing applications. With the human-centric design, SSEC envisions to integrate human intelligence into the process of data collection, processing, analysis, and decision making. We discuss several emerging applications that are enabled by SSEC, together with a number of open research challenges are to be undertaken by the community. We hope this paper will bring the SSEC paradigm to the attention of the community.